\documentclass[aps,prl,twocolumn,superscriptaddress]{revtex4}
\usepackage{graphicx}
\newcommand{\beq}{\begin{equation}}
\newcommand{\eeq}{\end{equation}}
\newcommand{\bea}{\begin{eqnarray}}
\newcommand{\eea}{\end{eqnarray}}
\newcommand{\pdag}{{\phantom{\dagger}}}

\begin{document}

\title{Frequency-dependent transport through a quantum dot in the Kondo regime}
\author{M.~Sindel}
\affiliation{Physics Department, Arnold Sommerfeld Center for Theoretical
Physics, and Center for NanoScience, Ludwig-Maximilians-Universit\"at
M\"unchen, 80333 M\"unchen, Germany}
\author{W.~Hofstetter}
\affiliation{Department of Physics, Massachusetts Institute of Technology, Cambridge MA 02139, USA}
\author{J.~von Delft}
\affiliation{Physics Department, Arnold Sommerfeld Center for Theoretical
Physics, and Center for NanoScience, Ludwig-Maximilians-Universit\"at
M\"unchen, 80333 M\"unchen, Germany}
\author{M.~Kindermann}
\affiliation{Department of Physics, Massachusetts Institute of Technology, Cambridge MA 02139, USA}

\date{\today}

\begin{abstract}
We study the AC conductance and equilibrium current fluctuations
of a Coulomb blockaded quantum dot in the Kondo regime. 
To this end we have developed  an extension of the numerical renormalization 
group suitable for the nonperturbative calculation of 
finite-frequency transport properties. 
We demonstrate that AC transport gives  access to the 
many-body resonance in the {\it equilibrium} spectral density. 
It provides a new route to measuring this key signature 
of Kondo physics which so far has defied direct
experimental observation.

\end{abstract}

\maketitle

In the past years semiconductor quantum dots have gained considerable
attention as tunable magnetic impurities \cite{Kondo-popular}. Due to
their small size, electronic transport through these structures is
strongly influenced by the Coulomb blockade \cite{Coulomb-blockade}.
 Quantum dots with an odd number of electrons display a well-known
 many-body phenomenon, the \emph{Kondo effect} \cite{Hewson,kondo}, as
 predicted in \cite{kondo-theo}. In these systems, a single unpaired spin 
is screened at low temperatures, giving rise to an enhanced DC 
(zero-frequency) conductance at low bias voltage.

Theoretical  studies have so far mostly focused on  DC
transport \cite{kondo-theo}.  From the differential conductance 
$dI/dV$ at finite bias
voltage $V$ one can obtain information about the Kondo resonance in the
spectral density. Such measurements on a typical, symmetrically coupled quantum dot, however, 
access only
the spectral function under non-equilibrium conditions.
It is known that in this case a finite bias splits and suppresses the Kondo peak \cite{Paa04}. 
It has also been demonstrated experimentally that a sufficiently strong AC modulation 
of the gate 
or bias voltage likewise suppresses the Kondo peak \cite{Elzerman00,Kogan04}.
Thus, it is by no means straightforward 
to measure 
the {\em equilibrium} spectral function of a Kondo quantum dot. 
Previous proposals to do so require 
either very asymmetric couplings to the leads or quantum dots in 
a 3-terminal geometry \cite{Sun01,Leb02}.

In this Letter we point out that the Kondo resonance 
in the equilibrium spectral density of a generic quantum dot
can be extracted from AC transport measurements.
We derive a direct relation between the spectral function and the 
linear AC conductance through a quantum dot [Eq.\ (\ref{eq:invert}) below]. 
Alternatively, the Kondo peak can be measured by
detecting equilibrium fluctuations in the current through the quantum
dot at frequencies of the order of the Kondo temperature $T_K$ [Eq.\ (\ref{eq:invertC}) below].  Such
measurements have come into experimental reach with recent advances in
the measurement of high-frequency current fluctuations \cite{Kou03}. 
Conductance measurements in the relevant frequency
regime are challenging, mainly due to parasitic currents through
hardly avoidable capacitances. One should, however, be able to extract
the signal due to the current through the quantum dot from this
background by an adiabatic change (much slower than the inverse Kondo 
temperature \cite{Nor99}) of gate voltage that drives the dot
periodically into and out of the Kondo regime.

We apply the nonperturbative 
numerical renormalization group (NRG) in combination with the 
Kubo formalism \cite{nrg,izumida97}
to predict the frequency-dependence of the AC conductance and the equilibrium
noise.
Our results are numerically essentially exact, in contrast to previous
perturbative calculations \cite{lopez01}, and 
generalize those of \cite{izumida97} to finite frequency. 
This allows us to access the 
experimentally most relevant regime of frequencies of the order of the Kondo
temperature for a device that exhibits a strong Kondo effect.

The exact correspondence between spectral function and AC conductance
holds only when charge fluctuations on the quantum dot are negligible.
Using our rigorous NRG results, we  verify that this condition is 
well satisfied for frequencies below the charging energy of the dot. 
\begin{figure}
\includegraphics[width=1.0\columnwidth]{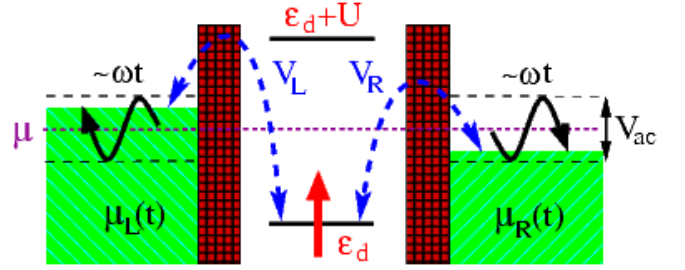}
\caption{Single-level Anderson model
with time dependent chemical potential 
$\mu_{\alpha}(t)=\mu \pm \left(e V_{\rm ac}/2\right) \cos \omega t $ in lead $\alpha \in \{L,R\}$.}
\label{fig:setup}
\end{figure}

The setup that we study is shown in Fig.~\ref{fig:setup}.  
At low energies, a Coulomb-blockaded quantum dot  is well described by the 
single-level Anderson model
\bea \label{And}
H &=& 
\sum_\sigma \epsilon_{d\sigma} 
d^\dagger_{\sigma} d^\pdag_{\sigma} + U n_{d \uparrow} n_{d \downarrow} +
\sum_{k \alpha \sigma} V_{\alpha} (d^\dagger_{\sigma} c^\pdag_{k\alpha\sigma} + h.c.)
\nonumber \\
&& + 
\sum_{k\sigma \atop \alpha \in \{L,R\}} \left[\epsilon_{k} - \mu_{\alpha}(t)\right] 
c^\dagger_{k \alpha \sigma} c^\pdag_{k \alpha \sigma},
\eea
where the chemical potentials 
$\mu_{L(R)}(t) = \mu \pm \left(e V_{\rm ac}/2\right) \cos \omega t $ 
for the left (right) lead include a time-dependent bias $V_{\rm ac}$. 
We  assume that the AC perturbation does not couple to 
electrons in the local level  $d^\pdag_{\sigma}$
($n_{d\sigma} = d^\dagger_{\sigma} d^\pdag_{\sigma}$).
$V_{\alpha} $ couples the  level $d$  
to electron states $c_{k\alpha\sigma}$ with momentum $k$ and spin $\sigma=\pm 1$
 in lead $\alpha \in \{L,R\}$. In the following, $D$ denotes the conduction electron bandwidth. 
 $\epsilon_{d\sigma}\equiv \epsilon_d + \sigma  B^* / 2$, $B^*=g\mu_B B$, 
is the energy of the local level, including the Zeeman shift 
in presence of a magnetic field $B$. $U$ is the Coulomb repulsion of  electrons on the dot.

Following \cite{jauho94} we relate the current $I$ through the quantum
dot to the local Green function of the level $d$.  For frequencies $\hbar\omega \ll \Delta_c$
much smaller than the charge excitation energy $\Delta_c = 
\min\{|\epsilon_{d}|,| U+\epsilon_{d}|\}$, charge fluctuations on the quantum
dot can be neglected and the currents flowing through the right and
the left lead are equal to a good approximation: $I_L=-I_R$. 
It is then
advantageous to write the total current as $I=(I_L-\lambda
I_R)/(1+\lambda)$, where $\lambda = \Gamma_L/\Gamma_R$,
$\Gamma_{\alpha}= \pi \nu V_{\alpha}^2$ and $\nu$ is the conduction
electron density of states.  Expressing the currents $I_{L}$ and $I_R$ by
the Green function of  $d$ (see Eq.\ (15) in \cite{jauho94})
and Fourier transforming from time $t$ to frequency $\bar{\omega}$, 
we find for
energy-independent couplings $\Gamma_{\alpha}$ the current 
\bea  \label{eq:curr}
I(\bar{\omega})&=& - \frac{2e^2 V_{\rm ac}}{\hbar^2 \omega} \frac{\lambda}{1+\lambda} 
\Gamma_R \int_{-\infty}^{\infty}{dt\, e^{i\bar{\omega} t} \int_{-\infty}^{\infty}
{dt_1 \int{\frac{d\epsilon}{2\pi} }}} \nonumber \\ 
&\times& {{{ {\rm Re} \,\left[ e^{i\epsilon(t-t_1)} 
(\sin\omega t-\sin\omega t_1) f(\epsilon) {\cal G}^r(t,t_1)\right]}}} 
\eea
in linear response to $V_{\rm ac}$.  
Here $f(\epsilon)$ is the equilibrium Fermi function of the leads 
and \mbox{${\cal G}^r(t_1,t_2)=-i \theta(t_1-t_2) \langle \{ d(t_1),d^\dag(t_2)\}\rangle$} 
the retarded Green function of $d$  evaluated in equilibrium ($V_{\rm ac}=0$).
From Eq.\
(\ref{eq:curr}) we obtain the real ($G'$) and imaginary ($G''$) parts of the
linear conductance, 
\bea \label{eq:MWcond} G'(\omega)
&=& \frac{e^2}{2\hbar^2 \omega} \frac{\Gamma_L \,\Gamma_R}{\Gamma}
 \int d\epsilon
f(\epsilon) \left[ A(\epsilon+\omega) - A(\epsilon-\omega) \right], \nonumber \\  \\ 
G''(\omega)
&=& \frac{e^2}{\hbar^2 \omega} 
\frac{\Gamma_L \,\Gamma_R}{\Gamma}
\int\frac{d\epsilon}{2\pi}
f(\epsilon) \Big[ 
-2 {\rm Re}\,{\cal G}^r(\epsilon)\nonumber \\ 
&& \;\;\;\;\;\;\;\;\;\;\;\;\;\;\; 
+ {\rm Re} \,{\cal G}^r(\epsilon+\omega) + {\rm Re}
\,{\cal G}^r(\epsilon-\omega) \Big] , \label{eq:MWcond_Im} 
\eea
where $\Gamma = \Gamma_L + \Gamma_R$ and  
$A(\omega)=-\frac{1}{\pi}{\rm Im}\, {\cal G}^r(\omega) $ is the spectral function.
At zero temperature $kT\ll \hbar |\omega|\ll \Delta_c$  Eq.\
(\ref{eq:MWcond}) is readily solved for the symmetrized spectral function  
$A_s(\epsilon)=[A(\epsilon)+A(-\epsilon)]/2$,
\beq \label{eq:invert}
A_s(\omega) =  \frac{\hbar^2}{e^2}\frac{\Gamma}{\Gamma_L \,\Gamma_R}
\frac{\partial}{\partial \omega} \left[ \omega \,
G'(\omega)\right].
\eeq
In the Kondo regime ($\langle n_{d}\rangle\simeq 1$), the Kondo resonance
is centered almost exactly around 0, thus the Kondo peaks in
$A_s(\omega)$ and $A(\omega)$ are essentially indistinguishable. 
Hence, Eq.\ (\ref{eq:invert}) allows us to extract the Kondo peak in the equilibrium
spectral function from a measurement of the AC conductance.
\begin{figure}
\includegraphics[width=1.0\columnwidth]{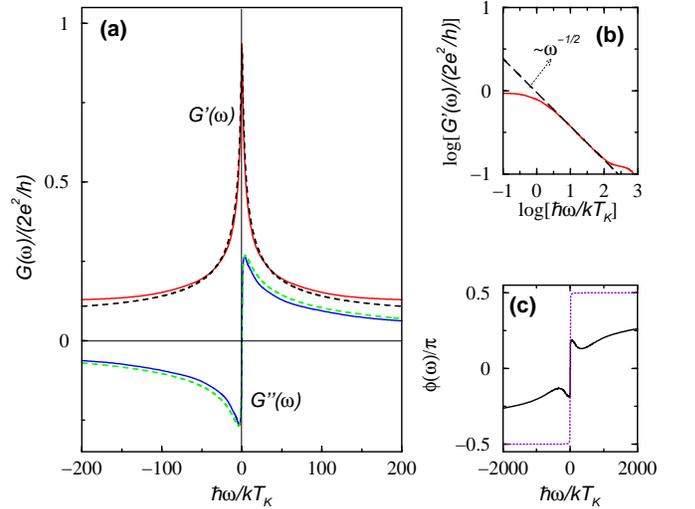}
\caption{(a) Real and imaginary parts of the AC conductance obtained via the 
Kubo approach (solid lines) and from the conductance formulas 
Eqs.~(\ref{eq:MWcond}) and (\ref{eq:MWcond_Im}) (dashed lines). 
Note the excellent agreement for frequencies $\hbar \omega \lesssim 100\,k\, T_K$
($\Delta_c \approx 600\,k\, T_K $).
(b) Doniach-Sunji\'c tails in $A(\omega)$ are responsible for the power-law behavior (dashed line) in the conductance (solid line) 
which is well described by $G' \propto \omega^{-1/2}$ for $20\,k\, T_K <\omega < 200\,k\, T_K$ \cite{doniach70}.
(c) Comparison of the frequency-dependent conductance phase $\phi(\omega) $ in the Kondo  
regime (solid line) (parameters $U=0.12D$, $U/\Gamma=6$, $\epsilon_d=-U/2$, $T=0$) 
with that for  a resonant level ($\epsilon_d=0$) of width $k\,T_K$ (dotted line).
}
\label{fig:G_omega}
\end{figure} 

The frequency-dependent equilibrium current fluctuations 
(Johnson-Nyquist noise)  
\beq \label{Nyquist}
C(\omega) = \int_{-\infty}^{\infty}
{dt\, e^{i\omega t}\,\left[\langle I(0) I(t)\rangle - \langle I\rangle^2\right]}
\eeq
are related to the linear conductance by the fluctuation dissipation theorem (FDT) 
\beq \label{eq:flucdiss}
C(\omega) =  \frac{2 \hbar \omega}{\exp(\hbar \omega/kT)-1}\, G'(\omega).
\eeq  
Consequently, the spectral function  can alternatively be inferred 
from  a measurement of $C(\omega)$. 
At zero temperature, $kT \ll \hbar |\omega|\ll\Delta_c$, we arrive at
 \beq \label{eq:invertC}
 A_s(|\omega|) = -\frac{\hbar}{2 e^2}
\frac{\Gamma}{\Gamma_L \,\Gamma_R} 
\frac{\partial}{\partial |\omega|} \, C(-|\omega|) .
\eeq  
Note that at zero temperature $C(\omega)$ is non-vanishing only for $\omega<0$, that is 
fluctuations have to be measured by probing absorption by the quantum dot.
In \cite{Kou03} the measurement of current fluctuations at frequencies
up to $\omega \simeq 100\, {\rm GHz}$ has been reported.  This frequency
scale is of the same order of magnitude as the Kondo temperature in
typical Kondo quantum dots \cite{Kondo-popular}. It  makes
 noise measurements  a promising candidate for experimental studies
of the Kondo peak in the spectral function of a quantum dot.

We turn now to a  discussion of  the frequency dependence of AC conductance 
and  equilibrium noise in a Kondo quantum dot with typical 
experimental parameters. We apply the numerical renormalization group (NRG),  
first used by Wilson to solve the Kondo problem \cite{nrg}. 
This method is nonperturbative and does not suffer from low-energy divergences 
common to scaling approaches. In particular, it provides accurate results 
in the most interesting crossover regime $\hbar \omega \lesssim k T_K$.
Within NRG we apply two independent approaches: 
Using the Kubo formula, we are able to calculate the conductance in the 
Anderson model Eq.\ (\ref{And})  numerically exactly. 
On the other hand, using Eqs.~(\ref{eq:MWcond}) and (\ref{eq:flucdiss}), 
we  obtain conductance and noise from the single-particle Green function of the 
$d$ level, that we  determine by NRG as well. By comparing the two approaches
 we shall demonstrate the validity of the approximation underlying Eqs.\
(\ref{eq:MWcond}) and (\ref{eq:MWcond_Im}) for frequencies of the order of the Kondo scale.

We apply the Kubo formalism following  Izumida \emph{et al.}~\cite{izumida97}.  
We define an electric current from lead $L$ to $R$ as 
$\hat{I} = \frac{e}{2} \left[\langle\dot{N_R}\rangle - \langle\dot{N_L}\rangle\right]$, 
where $N_{\alpha}=\sum_{k\sigma}
c^\dagger_{k \alpha \sigma} c^\pdag_{k \alpha \sigma}$ 
is the total number of electrons in lead $\alpha$
and $\dot{N_{\alpha}}=\frac{i}{\hbar}\left[H,N_{\alpha}\right]$.   
Introducing  a linear response tensor $\sigma$ by 
$\langle\dot{N}_\alpha\rangle = \sigma_{\alpha \beta}(\omega) \mu_{\beta}'$, 
where $\mu_{L(R)}'=\pm \left(e V_{\rm ac}/2\right) \cos \omega t $ 
is the time-dependent bias applied to the left (right) lead, 
the total linear AC conductance takes the form
\beq
G(\omega) = \frac{e^2}{4} \left[\sigma_{LL}(\omega) + \sigma_{RR}(\omega) 
- \sigma_{LR}(\omega) - \sigma_{RL}(\omega)\right]. 
\eeq
The complex response tensor $\sigma$ can be written as 
$\sigma_{\alpha \beta} (\omega) = \frac{1}{i \omega} 
\left[K_{\alpha \beta}(\omega) - K_{\alpha\beta}(0)\right]$, where 
\beq \label{response_tensor}
K_{\alpha\beta}(\omega) = -\frac{i}{\hbar} 
\int_0^\infty dt e^{-\delta t + i \omega t} 
\left\langle \left[\dot{N}_\beta(0), \dot{N}_\alpha (t)\right]\right\rangle  
\eeq
with $\delta\rightarrow 0^+$.
$\langle \ldots \rangle$ in Eq.~(\ref{response_tensor}) refers to the equilibrium
expectation value with respect to  the Hamiltonian $H$ with  $V_{\rm ac}=0$.
We evaluate this expression using NRG~\cite{nrg}.
From the matrix elements 
of the current operator $\langle n|\hat{I}|m \rangle$ the imaginary part 
$K_{\alpha\beta}''$ is obtained directly while the real part 
$K_{\alpha\beta}'$ can be calculated via 
a Kramers-Kronig transformation.

In Fig.~\ref{fig:G_omega} we compare the conductance obtained from 
Eq.\ (\ref{eq:MWcond}) with the result of the calculation in the Kubo formalism for $\epsilon_{d} = - U/2$
and $B=0$.  
We find excellent agreement for frequencies  below the charge gap 
$\Delta_c$. 
Moreover the large frequency ($\hbar \omega \gg k T_K$) asymptotes of the
conductance reveal a decay of $G'(\omega) \sim \omega^{-1/2}$ 
\mbox{[Fig.\ \ref{fig:G_omega} (b)]}. 
This is expected as a consequence of the Doniach-Sunji\'c tails of the spectral
 function \cite{doniach70} together with Eq.~(\ref{eq:invert}).
The deviation  of $G'(\omega)$ from  the unitary limit (by $\sim 6\%$) is due to 
systematic numerical errors accumulating in the NRG procedure.
 Fig.\ \ref{fig:G_omega}(c) shows the 
 frequency-dependent phase $\phi(\omega)$
 of the linear conductance, $G(\omega)=\left| G(\omega)\right| e^{i\phi(\omega)}$.

\begin{figure}
\includegraphics[width=1.0\columnwidth]{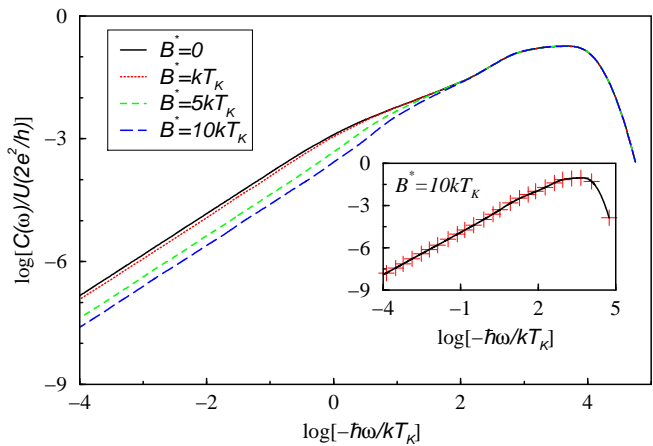}
\caption{
The equilibrium current fluctuations $C(\omega)$  
depend linearly on $\omega$ for $\hbar\left|\omega\right|\leq k T_K$~\cite{TKestimation},
as shown for various magnetic fields $B$.
The inset shows that the fluctuations are spin-independent 
even in presence of a finite magnetic field $B^*$ ($B^*=10\,k T_K$, (red) crosses: $\sigma=\uparrow$, 
(black) solid line: $\sigma=\downarrow$). 
Parameters: $U=0.12D$, $U/\Gamma=6$, $\epsilon_d=-U/2$, $T=0$.}
\label{fig:noise}
\end{figure} 
The frequency-dependence of the equilibrium noise obtained by a direct numerical evaluation of Eq.\ (\ref{Nyquist}) within NRG 
is plotted in Fig.~\ref{fig:noise}.
As one expects, the fluctuations reach a maximum for frequencies $\hbar\omega\sim U$.
The linear behavior $C(\omega) \sim \omega$ at small $\omega$ is a manifestation of the 
Fermi-liquid nature of a screened Kondo impurity. Fig.\ \ref{fig:noise} also shows 
how a finite magnetic field $B$  suppresses low-frequency equilibrium fluctuations. 
The inset demonstrates that at low frequencies the noise for 
spin up and spin down electrons is identical, even in the presence of a magnetic 
field. This is because an impurity with large charging energy 
$\Delta_c \gg \Gamma$ is half-filled: 
$\langle n_{d\uparrow} \rangle + \langle n_{d\downarrow} \rangle =1$. 
As a consequence of the Friedel sum rule~\cite{Langreth66}, the spectral densities
at the Fermi energy $\epsilon_F$ for spin-up and spin-down electrons 
$A_{\uparrow /\downarrow}(\epsilon_F) \propto 
\sin^2(\pi \langle n_{d\uparrow /\downarrow} \rangle)$  
are then equal.

\begin{figure}
\includegraphics[width=1.0\columnwidth]{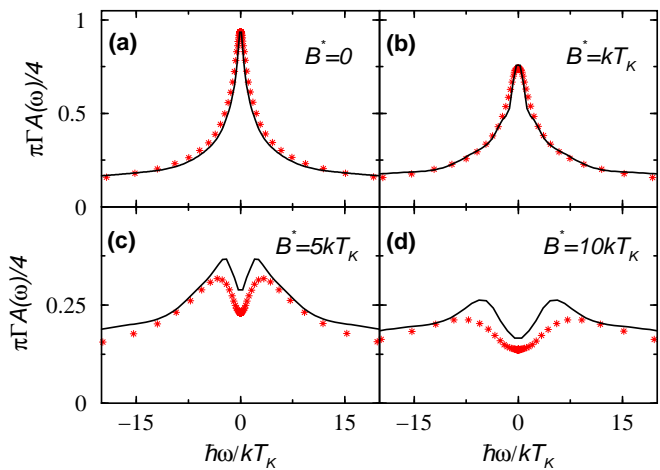}
\caption{Spectral function $A(\omega)\equiv A_s(\omega)$ as extracted from the current noise $C(\omega)$ via
Eq.~(\ref{eq:invertC}) (lines) compared to $A(\omega)$ obtained directly by NRG (stars) 
for different values of the  magnetic field $B$.
  Eq.~(\ref{eq:invertC}) captures the Kondo peak in   
 $A(\omega)$ very well, including  suppression and splitting in presence of 
a magnetic field $B$.}
\label{fig:spectrum}
\end{figure}

We now use Eq.\ (\ref{eq:invertC}) to extract 
the spectral function $A_s$ from the numerical current noise data of Fig.~\ref{fig:noise} where 
charge fluctuations on the dot have been taken into account. 
The results are shown in Fig.~\ref{fig:spectrum}. 
They nicely demonstrate the splitting and the suppression 
of the Kondo peak upon increasing the
external field $B$. A comparison with the spectral function directly
calculated by NRG confirms that Eq. (\ref{eq:invertC}) does indeed work
very well for frequencies $\hbar\omega\ll \Delta_c$.
In particular, the Kondo peak in the symmetrized spectral function
$A_s(\omega)$
obtained by our method is essentially indistinguishable from that in the
spectral function $A(\omega)$ directly calculated by NRG.
Due to intrinsic broadening in the NRG  
which affects both $A(\omega)$ and $C(\omega)$, 
Eq.~(\ref{eq:invertC}) becomes less accurate for finite $B$.

Having demonstrated that neglecting charge fluctuations in
deriving Eqs.~(\ref{eq:invert}) and
(\ref{eq:invertC}) is very well justified, we comment on another
limitation of our approach. Calculating the linear response of the
quantum dot, we have assumed the bias voltage in a conductance
measurement to be small. 
While this in itself is not a problem, one might
worry, however, that due to the finite frequency of $V_{ac}$
extra decoherence processes would make the limit of linear response
very restrictive.  The decoherence rate $\tau$ of the dot's spin due
to the oscillating bias voltage for $\hbar \omega \gg k\,T_K$ can be estimated
\cite{Kam00} as
\beq \label{eq:dec}
\frac{\hbar \tau^{-1}}{k T_K} \sim \left( \frac{e V_{\rm ac}}{k T_K}\right)^2 \, 
\frac{k T_K }{ \hbar \omega} \frac{1}{\left[\ln(\hbar\omega /k T_K)\right]^{2}}.
\eeq 
For the  Kondo physics not to be disrupted by $V_{\rm ac}$, we need 
$\hbar \tau^{-1}/k T_K\ll 1$.  Eq.\ (\ref{eq:dec}) shows that this
condition can be easily fulfilled by the usual requirement $e V_{\rm
ac}\ll k T_K$. We expect that this statement remains true for
frequencies of the order of $T_K$. We conclude that transport is
described accurately by our linear response conductance for voltages
that are much smaller than the Kondo temperature.

In conclusion, we have studied AC transport through a quantum dot in the 
Kondo regime, using the numerical renormalization group technique in combination 
with the Kubo formalism. 
We have expressed linear  conductance and equilibrium current fluctuations
  in terms of the single-particle Green function.  
This relation  becomes exact at low frequencies, when   charge fluctuations 
on the dot can be neglected. 
It has been   shown to work very well for frequencies of the order of the  Kondo scale. 
This opens up the exciting possibility of measuring the {\em equilibrium} Kondo resonance 
directly in a transport measurement. 

Acknowledgements: We would like to thank S.~Amasha, L.~Borda, M.~Kastner, and A.~Kogan
for valuable discussions, and L.~Glazman for  comments on the manuscript.
Financial support from SFB 631 of the DFG and CeNS is gratefully acknowledged.
W.H. has been supported by a Pappalardo grant;  
M.K. by the Cambridge-MIT Institute Ltd.


\begin{thebibliography}{99}
\bibitem{TKestimation} $T_K$ corresponds to the frequency at which  the spin spectral function 
takes its maximum.
\end{thebibliography}
\end{document}